\documentclass{Interspeech2024}
\usepackage{amsmath}
\usepackage{graphicx} 
\usepackage{subcaption} 
\usepackage{refcount}
\usepackage{multirow}
\usepackage{diagbox}
\interspeechcameraready

\title{HybridVC: Efficient Voice Style Conversion with Text and Audio Prompts}

\name[affiliation={1}]{Xinlei}{Niu}
\name[affiliation={1}]{Jing}{Zhang}
\name[affiliation={1}]{Charles Patrick}{Martin}

\address{
  $^1$Australian National University}
\email{xinlei.niu@anu.edu.au, zjnwpu@gmail.com, charles.martin@anu.edu.au}

\keywords{voice style, hybrid prompt, contrastive learning}

\begin{document}

\maketitle

\begin{abstract}

We introduce HybridVC, a voice conversion (VC) framework built upon a pre-trained conditional variational autoencoder (CVAE) that combines the strengths of a latent model with contrastive learning. HybridVC supports text and audio prompts, enabling more flexible voice style conversion. HybridVC models a latent distribution conditioned on speaker embeddings acquired by a pretrained speaker encoder and optimises style text embeddings to align with the speaker style information through contrastive learning in parallel. Therefore, HybridVC can be efficiently trained under limited computational resources. Our experiments demonstrate HybridVC's superior training efficiency and its capability for advanced multi-modal voice style conversion. This underscores its potential for widespread applications such as user-defined personalised voice in various social media platforms. A comprehensive ablation study further validates the effectiveness of our method.
\end{abstract}

\section{Introduction}
Voice conversion (VC) entails the process of modifying the vocal characteristics of a source speech to align with those of a target speaker. The key objective of VC is to alter the vocal identity, such as accent, tone, and timbre, to resemble the target speaker's voice style, while meticulously retaining the linguistic content of the source speech~\cite{mohammadi2017overview,zhou2021seen,wang2018style}. 


Traditional VC techniques rely on parallel training data~\cite{toda2007voice,desai2009voice,sun2015voice,zhang2019sequence},
which constrains their applicability for adapting to the voice styles of unseen speakers. 
To diversify the utilization of VC models, recent works focus on non-parallel any-to-any (A2A) VC models, so-called one-shot VC, that convert any speech to any voice style according to a few seconds of the target voice~\cite{qian2019autovc}. 
The key idea of A2A VC is learning disentangled feature representations of content and speaker style identity.
To disentangle content and speaker information, existing works focus on methods including instance normalization~\cite{park2023triaan},
vector quantization~\cite{wu2020one,wu2020vqvc+,wang2021vqmivc}, transfer learning from ASR and TTS models ~\cite{zhang2023leveraging,li2023styletts, hussain2023ace, rekimoto2023wesper}, and adversarial training~\cite{park2023triaan, li2023freevc,li2022asgan}.

With the development of large language models (LLMs), leveraging natural language prompts for generating images~\cite{gu2022vector}, sound~\cite{yang2023diffsound,liu2023audioldm,huang2023make,niusoundlocd}, and speech~\cite{wang2023neural,yang2023instructtts} has gained popularity. A notable development in VC is the introduction of voice style transfer using text prompts, as proposed by \cite{prompt_voice}. Building on this concept, \cite{kuan2023towards} introduced Text-guidedVC,
diverging from traditional VC techniques using natural language instructions, rather than an audio reference, to guide the VC process. This shift enhances the framework's flexibility and allows for a wider range of conversion styles by incorporating descriptors in natural language. Concurrently, PromptVC~\cite{yao2023promptvc} proposed an A2A text prompt-based VC model, following VITS~\cite{vits} structure, that employs a latent diffusion model (LDM)~\cite{LDM} to generate style vectors by text prompts.

We observe two limitations in Text-guidedVC~\cite{kuan2023towards}. Firstly, it relies on parallel training data, where effectiveness is heavily dependent on the size and quality of the dataset. This requirement can pose challenges, as accumulating extensive parallel data is often resource-intensive. Secondly, unlike conventional VC models that use audio reference, \cite{yao2023promptvc,kuan2023towards} may struggle with achieving precise voice conversions. In practical scenarios, text descriptions may fall short in accurately capturing and conveying specific voice styles, a gap that is readily bridged when using actual target voice samples. This limitation is crucial, as it affects the framework's ability to replicate exact voice styles based solely on text guidance. Therefore, an A2A VC model with hybrid prompts can be developed to combine the strengths of flexibility of text guidance and precision of voice guidance.

VC models~\cite{casanova2022yourtts,yao2023promptvc,li2023freevc} based on VITS~\cite{vits} framework may suffer from slow training due to decoding raw waveforms directly.
A straightforward solution is to avoid the decoding step during training but maintain its advantages on performance.
PromptVC~\cite{yao2023promptvc} trains an LDM jointly with a VITS framework to learn textual information. Although~\cite{yao2023promptvc} doesn't provide experimental setup details, jointly training an LDM and a CVAE with adversarial training requires a large volume of training data, powerful computational resources, and long training time that potentially limits its extension on practical applications. 
In image editing tasks, Imagic~\cite{kawar2023imagic} proposes text embedding optimisation,
which focuses on optimising text embeddings under pre-trained language models. This method offers a more flexible solution that avoids retraining the entire model, thereby providing an easier and more efficient way to extend the model through few-shot learning.

\begin{figure*}
    \vskip -0.2in
    \centering
    \includegraphics[width = 0.88\textwidth]{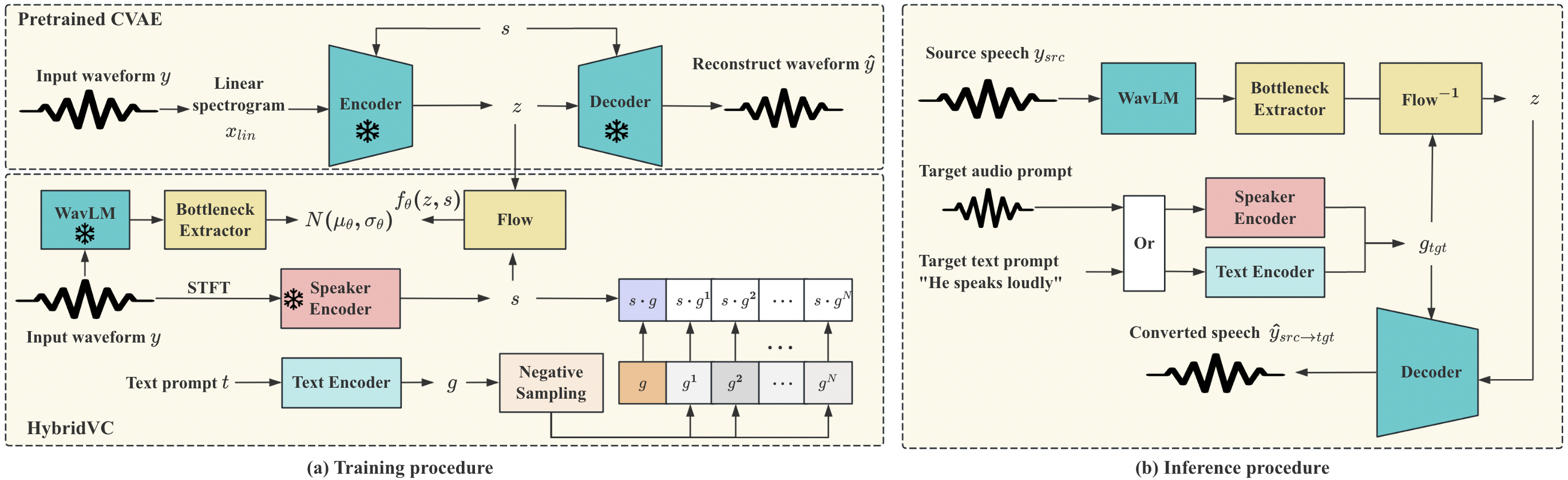}
    \vspace{-0.2cm}
    \caption{Overview of HybridVC, where $z$ represents the latent variable obtained by CVAE backbone, $g$ is text embeddings, $s$ is speaker style embeddings, and $\{g^1,...,g^N\}$ are $N$ negative samples. The snowflake symbol represents frozen parameters during training.}
    \label{fig:hybridvc-overview}
    \vskip -0.2in
\end{figure*}


In this work, we introduce HybridVC, an efficient A2A VC model that supports text prompts and audio prompts of target style voice to achieve better flexibility. 
HybridVC is a latent model with contrastive learning, 
which models the latent distribution conditioned on speaker style information and optimises the alignment of corresponding text embedding in parallel. 
Our experiments prove HybridVC achieves significant training efficiency under limited computational resources and achieves a flexible VC system that supports hybrid prompts. HybridVC supports small-scale training which can be easily adapted to applications such as user-defined personalised voice.

\section{Methodology}

\subsection{HybridVC}
HybridVC, shown in Figure~\ref{fig:hybridvc-overview}, is a conditional latent model that relies on a pre-trained CVAE backbone. 
We denote a text-speech pair $(y,t)$, where $y$ is an utterance and $t$ is the corresponding style text description. 
Similarly to LDM~\cite{LDM}, the first stage of HybridVC obtains a compact latent representation $z$ from the pre-trained CVAE conditioned on speaker style information $s$ by a pre-trained speaker encoder, as $z \sim q(z|x_{lin},s)$.
HybridVC~\footnote{Demonstration page: \url{https://xinleiniu.github.io/HybridVC-demo/}} models the distribution $p_\theta(z|y,s)$ conditioned on speaker embedding $s$ and raw waveform $y$. Following~\cite{li2023freevc}, the architecture of HybridVC is straightforward, using a pre-trained WavLM~\cite{chen2022wavlm} and a bottleneck extractor~\footnote{The bottleneck extractor is a projection layer used to reduce feature dimension and further disentangle content information $c$ from outputs of WavLM~\cite{chen2022wavlm}. We use the same setting as the one in~\cite{li2023freevc}.} to disentangle text-free content information $c$ from waveform $y$, a pre-trained speaker encoder to obtain speaker style information $s$, and a normalizing flow conditioned on $s$ to transform the distribution to a more complex distribution. This method enables HybridVC to encapsulate the content $c$ and speaker style information $s$ within a compact latent space, facilitating efficient manipulation and generation of speech.

To enable HybridVC to support audio and text prompts, the model incorporates a text encoder to optimise a text embedding $g$ through contrastive learning, as detailed in Section \ref{sec:ist}. The contrastive learning fine-tunes the text encoder, thereby, refining text embedding $g$ to be well-aligned with speaker style embedding $s$ obtained by the pre-trained speaker encoder.

\noindent\textbf{CVAE backbone}
The CVAE backbone compresses the input waveform $y$ into a latent variable $z$ conditioned on a speaker-style embedding $s$. In our experiments, we use the posterior encoder and decoder of a pre-trained FreeVC~\cite{li2023freevc} as the CVAE backbone which is augmented with GAN training.

\noindent\textbf{Speaker Encoder}
The speaker encoder extracts the vocal identity (i.e., speaker style embedding $s$) from $y$. We follow FreeVC~\cite{li2023freevc} that adopt a pre-trained verification model~\cite{liu2021any} as speaker encoder. The model is trained on a large amount of speakers. The speaker encoder embeds Mel-spectrogram of waveform $y$ into a speaker style embedding $s$, as $s\in \mathbb{R}^{256}$. 

\noindent\textbf{Text Encoder}
The text encoder extracts stylistic features from text prompt $t$. We fine-tune a pre-trained text encoder to refine the text embeddings $g$, ensuring they are closely aligned with the corresponding speaker style information $s$. The text encoder contains a RoBERTa~\cite{liu2019roberta} model from a pre-trained CLAP~\cite{clap} that encodes a text prompt $t$ into an embedding $g\in \mathbb{R}^{512}$, and a linear layer further compress $g$ to a lower feature dimension $g \in \mathbb{R}^{256}$ to discard irrelevant textual information.

\subsection{PromptSpeech + Text Embedding Optimisation}\label{sec:ist}

\begin{figure}
    \vskip -0.2in
    \centering
    \includegraphics[width=0.44\textwidth]{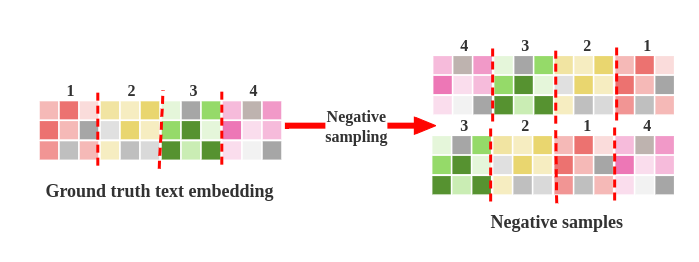}
    \vspace{-0.5cm}
    \caption{Illustration of negative sampling for text embedding.}
    \label{fig:negative_samples}
    \vskip -0.2in
\end{figure}

\begin{figure}
    \vskip -0.1in
    \centering
    \includegraphics[width=0.45\textwidth]{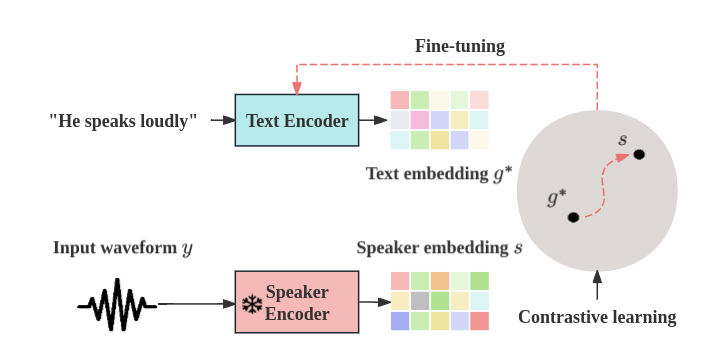}
    \vspace{-0.5cm}
    \caption{Illustration of optimising the text embedding $g^*$.} 
    \label{fig:embedding_optimization}
    \vskip -0.2in
\end{figure}

We use the PromptSpeech dataset~\cite{guo2023prompttts}, which pairs over 26,000 real speech samples from LibriTTS~\cite{zen2019libritts} with free-text style prompts involving categories of gender, pitch, energy, and speed. 
However, text style prompts in PromptSpeech do not guarantee comprehensive coverage of information across these four categories. 
For instance, a prompt like ``he said loudly'' specifies gender and energy but not pitch or speed. 
This inconsistency presents a challenge to ensure comprehensive representations of all stylistic attributes, particularly when attempting to precisely map these free-form textual prompts to a predefined set of labels for speaker embedding $s$.
To address this, we augment the data with $N$ negative samples on text embeddings $g$ extracted by the text encoder. Inspired by~\cite{cdcd,luddecke2022image}, we propose a negative sampling method for text embeddings $g$ that randomly shuffle $g$ into $N$ negative samples $G' = \{g^1,...,g^N\}$ as in Figure~\ref{fig:negative_samples}. Given the ground truth text embedding $g$, we construct $G'$ by dividing $g$ into $K$ randomly ordered chunks. 
As illustrated in Figure~\ref{fig:embedding_optimization},
HybridVC fine-tunes the text encoder through contrastive learning to optimise text embedding.

\begin{table*}[th]
  \vskip -0.1in
  \renewcommand{\arraystretch}{0.8}
  \renewcommand{\tabcolsep}{1.2mm}
  \caption{Model comparison measured on unseen source speeches with audio prompts from 9 unseen speakers.
  }
  \vskip -0.25in
  \label{tab:exp_overall}
  \begin{center}
  \begin{small}
  \begin{tabular}{c|cc|cc|cc|ccc}
    \toprule
    Model & Training Config. & Dataset & WER $\downarrow$ & CER $\downarrow$ & F$_0$-PCC $\uparrow$ & SSIM $\uparrow$ &FAD$\downarrow$ & FD $\downarrow$ & NISQA $\uparrow$\\
    \midrule
    FreeVC  & Nvidia RTX3090 & VCTK  &  13.56\% & 4.77\% & 0.719 & 0.781  & 1.15 & 0.71  & 4.51\\
    FreeVC $\star$& 2 Nvidia RTX3090 (12 days) & VCTK &  \textbf{13.16}\% & \textbf{4.60}\% & \textbf{0.720}  & 0.771& 0.89 & 0.71 & 4.34 \\
    YourTTS-VC  & Nvidia TESLA V100 64G & VCTK & 28.50\% & 12.45\% & 0.637 & 0.662  & 3.99 & 1.37 & 3.88 \\
    HybridVC $\diamond$& 2 Nvidia RTX3090 (15 hours) & VCTK  &  14.01\% & 4.85\% & 0.716 & \textbf{0.783}  & 0.87 & \textbf{0.70} & 4.51 \\ 
    \midrule
    HybridVC $\diamond$ & 2 Nvidia RTX3090 (17 hours)& PromptSpeech &  18.18\% & 6.15\% & 0.701& 0.771 &  \textbf{0.74} & 0.81 & 4.51 \\
    HybridVC & 2 Nvidia RTX3090 (25 hours)& PromptSpeech &  17.54\% & 6.17\% & 0.712& 0.775  & 0.79 & 0.75  & \textbf{4.52} \\
    \bottomrule
  \end{tabular}
  \end{small}
  \end{center}
  \vskip -0.3in
\end{table*}

\subsection{Training}
As in Figure~\ref{fig:hybridvc-overview}(a), the loss function of HybridVC contains two parts: a KL-divergence between the latent distributions of HybridVC and CVAE, and a contrastive loss between text embedding with negative samples $\{g, G'\}$ and speaker embedding $s$.
Following~\cite{vits}, the KL loss $\mathcal{L}_{kl}$ between a Gaussian and normalising flow optimises the latent distribution of HybridVC,
\begin{align}
    \mathcal{L}_{kl}  & = \text{log}q(z|x_{lin},s)-\text{log}p_\theta(z|y,s) \label{eqn:kl}
\end{align}
where $f$ is the normalising flow, $x_{lin}$ is the linear spectrogram, 
\begin{align}
    p_\theta(z|y,s) & =N(f_\theta(z,s);\mu_\theta,\sigma_\theta)|\frac{\partial f_\theta(z,s)}{\partial z}| \\
    q(z|x_{lin},s) & = N(z;\mu_q,\sigma_q)
\end{align}
The contrastive loss $\mathcal{L}_{contrastive}$ is used to optimise the text encoder aligning the text embedding $g$ with the speaker embedding $s$. Given a set of negative sample $G' = \{g^1, ..., g^N\}$, the contrastive loss is defined as
\begin{equation}
    \mathcal{L}_{contrastive} =\text{log} \frac{exp(cos(g,s)/\tau)}{\sum_{g_i \in \{g,G'\}} exp(cos(g_i,s)/\tau)}
\end{equation}
where $\tau$ is a learnable temperature parameter.

The overall loss function $\mathcal{L}$ in HybridVC is defined as 
\begin{equation}
    \mathcal{L} =  (1-\alpha)\mathcal{L}_{kl} + \alpha \mathcal{L}_{contrastive}
\end{equation}
where $\alpha$ is a weight-scaling hyper-parameter parameter.

\subsection{Inference}

As in Figure~\ref{fig:hybridvc-overview}(b), given a source speech $y_{src}$, and a prompt embedding $g_{tgt}$ obtained from either the speaker encoder or the text encoder, the inference phase of HybridVC is
\begin{equation}
    \hat{y}_{src \rightarrow tgt} = Decoder(z, g_{tgt})
\end{equation}
where $z \sim p_\theta(z|y_{src},g_{tgt})$.

\section{Experiments and Results}

\noindent\textbf{Datasets}
We conduct experiments on VCTK~\cite{vctk} and PromptSpeech~\cite{guo2023prompttts} datasets. PromptSpeech includes over 26000 real utterances from LibriTTS~\cite{zen2019libritts} with corresponding text style prompts, it is used for training HybridVC only. For VCTK, we follow the train/test separation as in~\cite{li2023freevc} which involves 107 speakers in total, 314 utterances for validation, 1070 utterances for testing and the rest for training\footnote{\url{https://github.com/OlaWod/FreeVC}}\label{fn:freevc_git}. VCTK is used to train HybridVC for a model comparison experiment. The VCTK testing set is used to verify models' performance as source speeches.

\noindent\textbf{Experimental setups}
All the audio samples are downsampled to 16kHz. The short-term Fourier transform (STFT) with a 1280 window size and 25\% overlapping is performed to extract linear and 80-bin Mel-spectrograms. 
Following~\cite{li2023freevc}, we use SR-based data augmentation during training, where the resize ratio $r$ ranges from 0.85 to 1.15. 
A HiFi-GAN v1 vocoder~\cite{kong2020hifi} is used for the data augmentation. 
The negative sample size $N$ is 10 and the number of shuffle chunks $K$ is 8. 
We use the posterior encoder and decoder of a FreeVC~\cite{li2023freevc} trained on the VCTK training set as the CVAE backbone.
All experiments were performed on 2 Nvidia RTX3090 GPUs. 
The model is trained with up to 80k training steps, where the batch size is 256 and initial learning rate is 2e-4.
The RoBERTa model is initialized by weights from a CLAP checkpoint~\footnote{\url{https://github.com/LAION-AI/CLAP}}.

\noindent\textbf{Models for comparison}
Since HybridVC is proposed to improve the flexibility and training efficiency compared to the VITS-like VC models, we select the following models for comparison\footnote{Since the closest methods mentioned previously, Text-guidedVC~\cite{kuan2023towards} and PromptVC~\cite{yao2023promptvc}, did not release implementation code and training datasets, we cannot use them for direct comparison.}:
1) FreeVC~\cite{li2023freevc}, an official checkpoint trained with SR data argumentation and a pre-trained speaker encoder on the VCTK training set, 2) FreeVC$\star$: a retrained 900k steps FreeVC with SR data argumentation and a pre-trained speaker encoder on the VCTK training set under the official training configuration\footnotemark[1], 3) YourTTS-VC~\cite{casanova2022yourtts}\footnote{\url{https://github.com/Edresson/YourTTS}}, officially released checkpoint on VCTK trained with pre-trained speaker encoder. 

Two versions of the proposed models are tested:
1) HybridVC$\diamond$, the proposed model trained with KL loss $\mathcal{L}_{kl}$, and 2) HybridVC, the proposed model trained with the total loss $\mathcal{L}$.

\noindent\textbf{Evaluation metrics}
We use nine objective evaluation metrics to verify the proposed method in diverse aspects including voice similarity, content error rates, speech naturalness, audio quality, and style prompt consistency. 
We calculate \textit{word error rate} (WER) and \textit{character error rate} (CER) between source and converted voice by an ASR model~\footnote{\url{https://huggingface.co/facebook/hubert-large-ls960-ft}}. We verify the voice similarity by \textit{Pearson correlation coefficient of fundamental frequency} (F$_0$-PCC) between source and converted voice, \textit{structural similarity index measure} (SSIM) between target and converted voice. F$_0$-PCC measures the similarity of pitch contours between converted voices and source voices. We then make use of \textit{speech quality and naturalness assessment} (NISQA)~\cite{mittag2021nisqa}, \textit{Fréchet audio distance} (FAD)~\cite{liu2023audioldm} and \textit{Fréchet distance} (FD)~\cite{liu2023audioldm} to verify the naturalness of converted voice and audio quality. The NISQA is a deep learning framework to predict speech quality and naturalness~\cite{mittag2021nisqa}, and FAD/FD measures audio quality. We calculate the cosine similarity (COS) between text embeddings extracted by the text encoder and audio embeddings of converted speeches extracted by the speaker encoder to verify the voice style consistency. Referring to \cite{kuan2023towards}, we calculate the classification accuracy between converted voice and source voice given single-factor text prompts. 

\subsection{Speech Intelligibility, Naturalness and Audio Quality}

Following experimental design in~\cite{li2023freevc}, we randomly select 9 unseen speakers as target voices and the test set of VCTK as source speeches. 
We aim to verify whether HybridVC improves training efficiency and maintains similar performance on audio prompt voice style conversion compared to baseline models, which only support audio prompt voice conversion.
We train HybridVC$\diamond$ and FreeVC$\star$ on the VCTK training set under the same configuration. 
Table~\ref{tab:exp_overall} shows that HybridVC can achieve competitive performance on speech intelligibility, naturalness, and audio quality with only 15 hours of training on limited computational resources.
Extending the experiment to include training on the PromptSpeech dataset, HybridVC maintains overall performance without noticeable degradation, despite the backbone CVAE only being pre-trained on the VCTK training set. This underscores the robustness of HybridVC and highlights its capability to adapt to new datasets with minimal training resources and time, further explaining the advantage of HybridVC on extensions of practical application.

\subsection{Consistency Test}

We then conduct a consistency test between converted voice and style prompts for both audio and text prompts. We randomly select 9 unseen speakers as audio prompts and the VCTK testing set as source speeches to test consistency between converted voices and style audio prompts. 
To test the text-style voice conversion consistency, we randomly selected 10 speakers (5 female and 5 male) including 5 utterances for each speaker as source speech and 5 style text prompts from the PromptSpeech.
As in Table~\ref{tab:prompt_consistency}, HybridVC effectively maintains the prosody of source speech and audio quality but accurately converts the voice characteristics given audio and text prompts. Text prompts typically offer less information compared to audio prompts, resulting in a slight drop in F$_0$-PCC and FAD values.
The COS between converted voice and text prompt, and SSIM between converted voice and audio prompt indicate the style of the converted voice is consistent with the provided prompts.

\begin{table}[t]
  \vskip -0.2in
  \renewcommand{\arraystretch}{0.9}
  \renewcommand{\tabcolsep}{1.5mm}
    \caption{Consistency test for audio prompts and text prompts}
   \vskip -0.1in
    \centering
    \begin{tabular}{ccccc}
        \toprule
         & F$_0$-PCC $\uparrow$ & FAD $\downarrow$ & SSIM $\uparrow$ & COS $\uparrow$\\
        \midrule
        Text prompts&  0.69 & 0.79 &  \diagbox{}{} & 0.68 \\ 
        Audio prompts & 0.71 & 0.76  & 0.78 & \diagbox{}{}  \\
        \bottomrule
    \end{tabular}
    \label{tab:prompt_consistency}
\end{table}

\begin{figure}
    \vskip -0.2in
    \centering
    \includegraphics[width=0.45\textwidth]{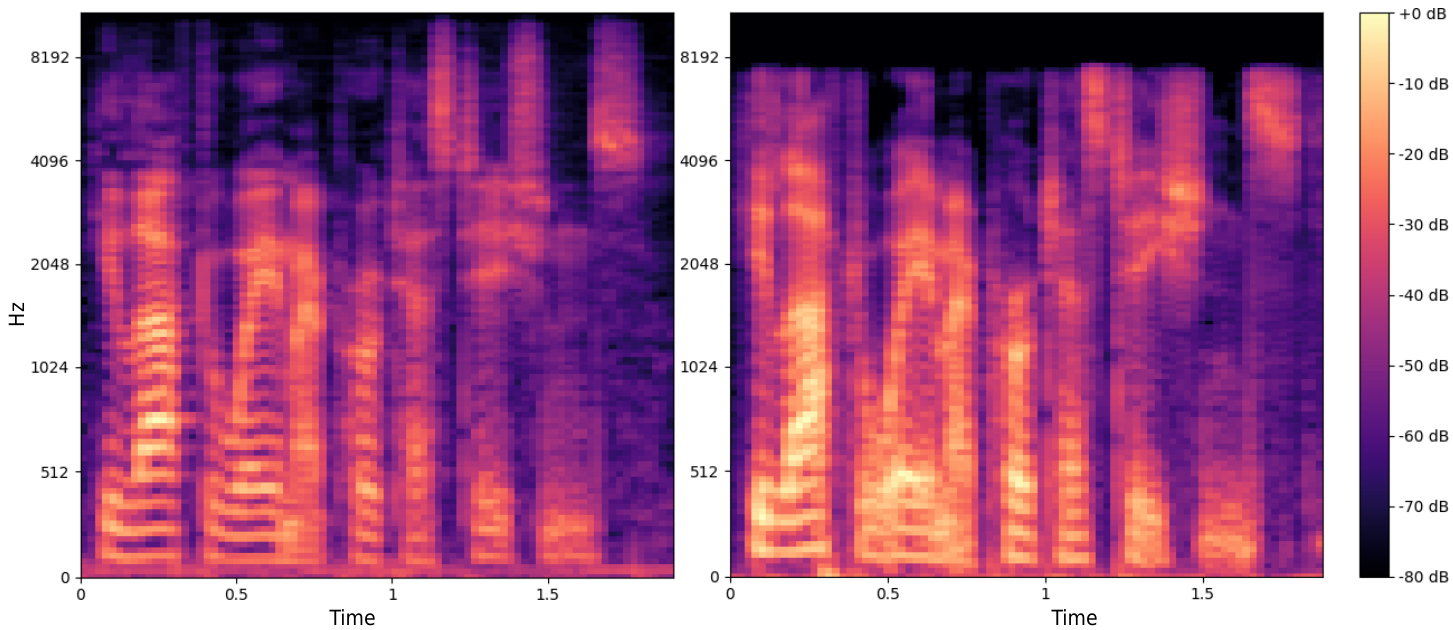}
    \vspace{-0.2cm}
    \caption{Visualization of source speech (left) and converted speech with text prompt ``he speaks loudly" (right).}
    \label{fig:vis}
\end{figure}

We then explore the congruence between source and converted voices using single-factor text prompts. 
Figure~\ref{fig:vis} provides an example of a visualization comparison of a converted voice by a single-factor text prompt.
In Table~\ref{tab:acc}, we calculate the accuracy percentage between the absolute average value of the converted voices and the source voices according to the text prompts. For instance, we count the number of converted voices with higher pitch under the ``higher pitch'' text prompt by comparing the average pitch value between converted voices and source voices.
Despite these prompts containing only single style factors, HybridVC successfully adapts voices to match the specified style text prompts. 
However, we observe that HybridVC demonstrates less sensitivity to text prompts that include words such as ``lower'', leading to a future improvement.

\begin{table}[t]
  \renewcommand{\arraystretch}{0.9}
    \vskip -0.1in
    \centering
    \caption{Accurcy of converted speech by given text prompts}
    \vskip -0.1in
    \begin{tabular}{ccccc}
        \toprule
        \multirow{2}{*}{Text prompt} &  Higher & Higher & Lower&Lower \\
        & pitch & volume & pitch & volume\\
        \midrule
        Accuracy &  89.8 \% & 91.1\% & 60.5\%  &  58.9\% \\  
        \bottomrule
    \end{tabular}
    \label{tab:acc}
    \vskip -0.2in
\end{table}

\begin{figure}
    \centering
    \vskip -0.2in
    \includegraphics[width=0.49\textwidth]{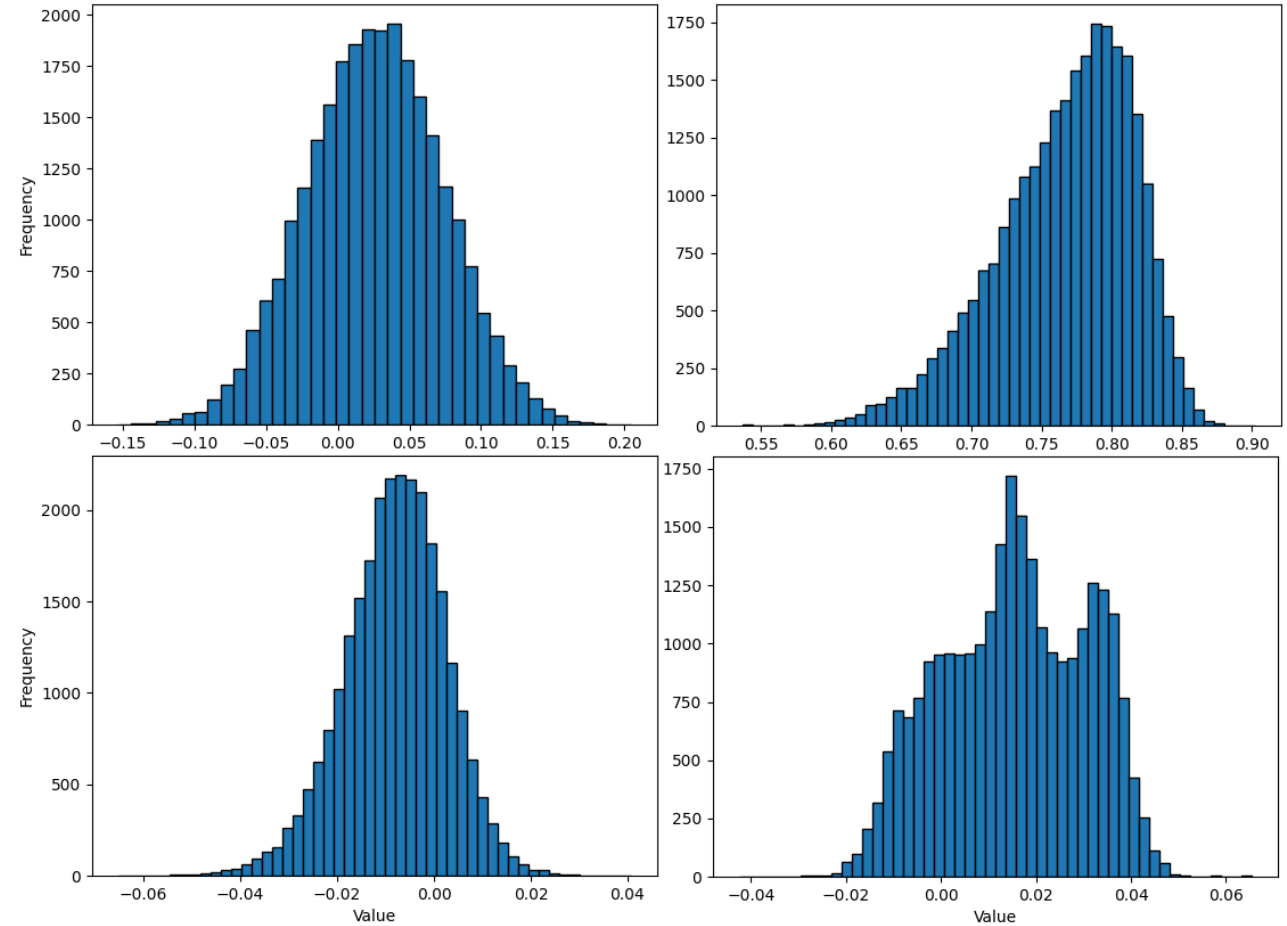}
    \vspace{-0.5cm}
    \caption{Cosine similarity distribution before text embedding optimisation (left-top); after fine-tuned with proposed negative sampling method (right-top); after fine-tuned with negative samples within the batch (left-bottom); after fine-tuned with negative samples with manually classified labels (right-bottom).}
    \vskip -0.1in
    \label{fig:ablation_cos}
\end{figure}

\subsection{Ablation Study}

\noindent\textbf{Latent model} 
We conduct an ablation study to verify whether the proposed latent model improves training efficiency while maintaining a similar performance as the baseline~\cite{li2023freevc}. 
As shown in Table~\ref{tab:exp_overall}, HybridVC$\diamond$ has great training efficiency, achieving competitive performance in only 15 hours of training. Adding the negative sampling method to HybridVC increases the training time while maintaining HybridVC's performance. This presents a reasonable trade-off between reducing training efficiency and enabling support for flexible hybrid prompts.

\noindent\textbf{Negative sampling method} To verify the proposed negative sampling method, we train the HybridVC augmented with: 1) construct negative samples within a training batch as in~\cite{clap}; 2) classify text embeddings into 54 labels across four categories in PromptSpeech. Figure~\ref{fig:ablation_cos} shows the cosine similarity between text embedding and speaker embedding before and after optimisation with different negative samples. Only HybridVC trained with the proposed negative sampling method effectively refines text embeddings well-aligned to speaker embeddings.

\begin{figure}
    \centering
    \includegraphics[width=0.33\textwidth]{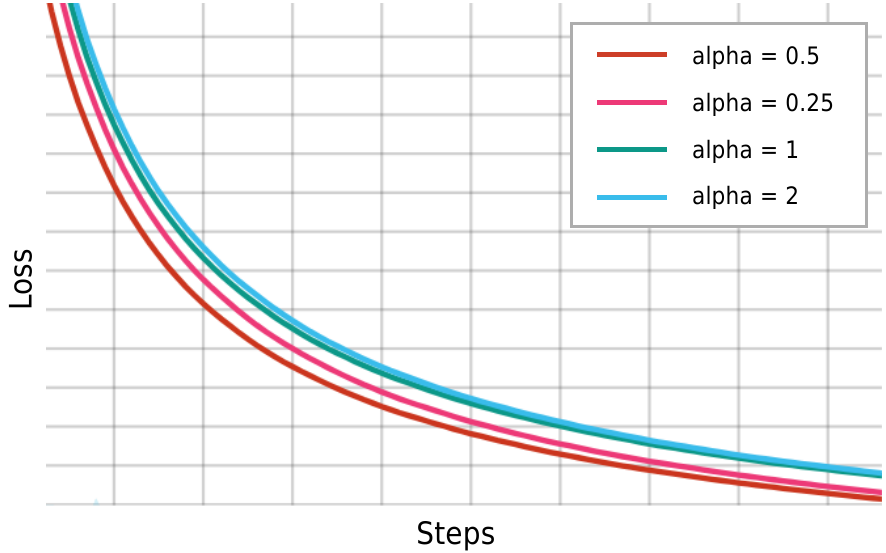}
    \vspace{-0.3cm}
    \caption{Convergence analysis plot against $\alpha$.}
    \label{fig:convergence}
    \vskip -0.25in
\end{figure}

\noindent\textbf{Hyper-parameter $\alpha$}
We conduct convergence analysis against $\alpha$ values. Figure~\ref{fig:convergence} shows that the loss convergence is not sensitive to $\alpha$ values although smaller $\alpha$ values (greater weighting on $\mathcal{L}_{kl}$) lead to slightly faster convergence.

\section{Conclusion}

We propose HybridVC, a flexible VC model supporting audio and text prompts that can be efficiently trained under limited computational resources. 
HybridVC is a latent model that 
models the latent distribution of a pre-trained CVAE backbone conditioned on speaker embeddings, and refines the text embedding to better align with information of speaker style in parallel. 
Our experiments show that HybridVC can achieve great training efficiency while maintaining robustness, flexibility, and performance compared to baseline models. 
Our model could be extended easily to applications, such as personalised voice on social media.
However, as the first VC model supporting hybrid text/audio prompts, one limitation is that the optimised text embedding appears to be less sensitive to text prompts including words such as ``lower''. 
Future work could improve text-speaker embedding alignments by techniques such as prompt tuning.
\newpage
\bibliographystyle{IEEEtran}
\bibliography{mybib}

\end{document}